\documentclass{jaa}

\usepackage{graphicx}

\usepackage{aas_macros} 
\usepackage{natbib}
\begin{document}

\title{Radio afterglows of Gamma Ray bursts\\}


\author{Lekshmi Resmi}
\affilTwo{\textsuperscript{1}Indian Institute of Space Science and Technology, Trivandrum - 695 547, India.}


\twocolumn[{

\maketitle

\corres{l.resmi@iist.ac.in}

\msinfo{6 May 2017}{xx}{yy}

\begin{abstract}
This review focuses on the physics of Gamma Ray Bursts probed through their radio afterglow emission. Even though radio band is the least explored of the afterglow spectrum, it has played an important role in the progress of GRB physics, specifically in confirming the hypothesized relativistic effects. Currently radio astronomy is in the beginning of a revolution. The high sensitive Square Kilometer Array (SKA) is being planned, its precursors and pathfinders are about to be operational, and several existing instruments are undergoing upgradation. Thus, the afterglow detection statistics and results from follow up programs are expected to improve in the coming years. We list a few avenues unique to radio band which if explored to full potential have the promise to greatly contribute to the future of GRB physics. 
\end{abstract}


}]


\doinum{10.1007s12036-017-9472-7}
\artcitid{1}
\volnum{1}
\year{2017}
\pgrange{1-14}
\setcounter{page}{1}
\lp{1}

\section{Introduction}
Gamma Ray Bursts (GRBs) were serendipitously discovered in late 1960s by the Vela military satellites. 
In the following years, dedicated scanning instruments on board high energy missions like \emph{BeppoSAX}\footnote{https://heasarc.gsfc.nasa.gov/docs/sax/sax.html}, \emph{CGRO}\footnote{https://heasarc.gsfc.nasa.gov/docs/cgro/cgro.html}, \emph{HETE}\footnote{
https://heasarc.gsfc.nasa.gov/docs/hete2/hete2.html}, \emph{Swift}\footnote{https://swift.gsfc.nasa.gov/} and \emph{Fermi}\footnote{https://fermi.gsfc.nasa.gov/} have increased the number of GRB detections to several thousands. GRBs are non-recurring events, hinting at underlying catastrophic phenomena.  The gamma-ray flash typically lasts for a few seconds to a few minutes, and in some rare cases to thousands of seconds. 

%
Most GRB detections in the initial decades were made by the \emph{BATSE} instrument on-board \emph{CGRO}. Due to the poor sky localization capacity of \emph{BATSE}, of the order of a few degrees, which is larger than the field of view of typical optical and radio telescopes, scope of longer wavelength follow up observations were limited in this period. Understanding the nature of these sources from unpredictable and short lived $\gamma$-ray emission alone was difficult. 
By the launch of \emph{BeppoSAX} in 1996 with a Wide Field Camera (WFC) capable of better angular resolution ($\sim 5$ arc minute) the era of long wavelength counterparts  began \cite{Cota:1997cg}.

The long wavelength ``afterglow emission'' lasts for longer time-scales, a few days in X-rays/Optical to even years in the radio bands. From its beginning two decades ago, afterglow studies have played a pivotal role in the progress of GRB physics. The first milestone was in confirming their cosmological origin \cite{vanParadijs:1997wr}. 

In most cases cosmological redshift ($z$) is measured from absorption lines against the bright optical afterglow of a burst or from emission lines from the associated host galaxy \cite{Galama:1997a}. As of now, the closest known burst is $\sim 39$~Mpc away ($z = 0.008$) and the furthest detected one till date is at a redshift of $9.4$ (photometric $z$, \cite{Cucchiara:2011pj}). GRBs are perhaps the cleanest known beacons of the early universe.

In this article, we review the progress made in GRB physics through radio afterglow studies. As of now, radio is the least explored band in the afterglow spectrum, a major reason being the poorer flux sensitivity of the radio instruments compared to Swift XRT or a $1$ to $2$~m class ground based optical telescope. With around $10^4$ seconds of integration, XRT sensitivity reaches around $4 \times 10^{-14}$erg cm$^{-2}$ s$^{-1}$ (reference: Swift XRT instrument page), which after roughly accounting for the bandwidth corresponds to $\sim 10^{-7}$~mJy. A limiting optical magnitude of $23$ in AB system corresponds to $\sim 1\, \mu$~Jy. For comparison, the Very Large Array (VLA) afterglow upper limits before upgradation was $\sim 100$~$\mu$ Jy. However, the Jansky VLA reaches to a few $10$~s of $\mu$ Jy sensitivity with around one hour on source time (ref: several GCN circulars). Therefore, number of radio detections and long follow-up campaigns have already started to increase. In a few years, the high sensitive Square Kilometer Array (SKA) will become operational. Its precursors and pathfinders are starting to be operational, and several existing instruments like the Giant Meter-wave Radio Telescope (GMRT) are undergoing upgradation. In this background, we list a few unique features of afterglow physics radio band alone can probe. There are two good reviews already existing with a similar theme. \cite{granotalexander} provides an extensive review of radio afterglows covering several theoretical and observational prospects. \cite{Chandra:2016tzf} reviews radio afterglows with more observational emphasis than us. We focus more on the theoretical afterglow modelling aspects and take a pedagogical approach throughout.

The review is arranged as follows. In the next section, we will give a brief description of some of the basic concepts in afterglow physics. In section-\ref{class}, we discuss the burst classification and progenitor models. Section-\ref{std} gives an overview of the standard afterglow model. We discuss the most important inferences that could be drawn from the radio band, especially with the support of broad-band modelling, in section-\ref{prob}. We devote section-\ref{q} for recent novel conjectures on potential diversity in radio afterglows. We conclude with a note on how upcoming facilities in radio band will be of importance in GRB physics.
\section{Basic energetics}
\label{basics}
The first step towards building a physical model for GRBs is to form an idea on the energetics involved. Once the luminosity distance $d_L$  is known from $z$, the energy can be estimated from the observed $\gamma$-ray fluence \footnote{Fluence is energy received per unit area.} ($f_{\gamma}$).  Assuming the energy is released isotropically, $E_{\gamma, {\rm iso}} = 4 \pi d_L^2 f_{\gamma}$.  However, the true energy budget will be different if the energy is released anisotropically. Typical values of $E_{\gamma, {\rm iso}}$ varies from $10^{50}$ to $10^{52}$ ergs. In at least one exceptional case, the isotropic energy budget has exceeded the rest energy of a solar mass object (GRB 080916C \cite{Abdo:2009zza}). 

The initial $\gamma$-ray flash, also known as prompt emission, is found to vary in milli second time-scale. This indicates, from the light travel time effects, that the emission region is compact ($R < c \delta t$ where $R$ is the size of the region, $c$ is the speed of light, and $\delta t$ is the variability time-scale of the prompt emission), and is of the order of $10^7$~cm. %
The compactness of the emission region, the energy involved in the process, and the typical local rate of $\sim 1 /Gpc^{-3} yr^{-1}$ \cite{Kumar:2014upa} indicate that Gamma Ray Bursts may originate in the gravitational collapse leading to the formation of stellar mass black-holes. 

Confining MeV gamma-rays in a compact volume of $\sim 10^{22}$~cm$^3$ will lead to a plasma that is optically thick to pair production, which in turn will produce a thermal spectrum in soft gamma-rays. However, the prompt emission spectrum is predominantly non-thermal. This paradox, known as the \textit{compactness problem}, could only be resolved by invoking a relativistically expanding source, a hypothesis proved right through radio afterglow observations (see section-\ref{scintil}).

\section{Burst classification and progenitor models}
\label{class}
During the \emph{BATSE} period itself, GRBs were found to fall in a bimodal distribution in the spectral and temporal characteristics of the prompt emission: short duration bursts with relatively harder $\gamma$-ray spectrum and  longer duration ones with a softer spectrum \cite{Kouveliotou:1993yx}. In the BATSE band, the long-short distinction is set at a $T_{90}$ of $2$~seconds. $T_{90}$ of a burst is the duration in which $5$\% to $90$\% of the total counts are emitted. However, since the burst duration is a function of the energy band in which it is observed, this value is detector dependent. The observed duration is also a function of redshift due to cosmological time dilation, and an intrinsically short burst  at high $z$ can be mis-identified as a long burst \cite{Zhang:2009uf}. 
Recently, a new class of bursts known as ultra-long bursts, which have thousands of seconds of duration, have been discovered \cite{Levan:2013gcz}. Though there are attempts to bring in a physically motivated classification scheme \cite{Zhang:2006mb}, by and large the long-soft and short-hard classification is followed.

In both long and short bursts, almost similar amount of energy (within two or three orders of magnitude) is released in a $\gamma$-ray flash, laden with milli-second scale random variabilities. The outflow that generates both kind of flashes has to be relativistic (see compactness problem mentioned in the previous section), which means the energy per baryon mass ($E/M_{\rm ej} c^2) \gg 1$. Yet, there is a distinction in the duration of prompt emission between long and short bursts. This indicates that while there are similarities in the way the $\gamma$-ray flashes are produced, the progenitors are likely to be different between the two classes.

Hence the current understanding is that the central engine that powers both the long and short bursts could be similar, but they might originate from two different channels. A black-hole torus system \cite{Woosley:1993wj} or a milli-second magnetar \cite{Usov:1992zd} are proposed for the central engine - the only two possibilities where $10^{50}$ to $10^{52}$~ergs of energy could be efficiently extracted. Since the burst duration is an indication of the life time of the jet emanating from the central engine,  long bursts need to have a more massive torus.  Hence long bursts are suggested to be due to the gravitational collapse of a massive star ($M > 15M_{\odot}$)  to a black hole \cite{Woosley:1993wj} while the short ones are believed to be originating from the merger of double neutron star binaries or neutron star - black hole (NS-BH) binaries \cite{Eichler:1989ve}. 

Further observations of the afterglow and host galaxies have provided supporting evidences for the above progenitor hypothesis. In a few nearby long bursts, the featureless blue continuum of the optical afterglow emission was found to give way to the redder supernova spectrum (stripped envelope type-Ic) rich with emission lines \cite{Galama:1998ea, Hjorth:2003jt}. This association of long GRBs with core collapse supernovae is the strongest support for the massive star progenitor model. Multi-messenger astronomy with gravitational wave and electromagnetic detectors is expected to soon unravel the association of short bursts and compact object mergers.

\section{Standard fireball model: fundamental layout}
\label{std}
Years of multi-wavelength observations and theoretical developments have laid a solid foundation of the standard fireball model, according to which the Gamma Ray Burst originates in a collimated relativistic outflow (jet) launched by a central engine: either a black-hole torus system or a fast rotating highly magnetized neutron star (milli-second magnetar). Unlike outflows which are imaged (for example, AGN jets or protostellar jets) collimation is an indirect inference for GRBs (see section-\ref{jet}). 

The jet energy is dissipated and radiated away in two phases. An internal dissipation either through shocks or magnetic reconnection leads to the prompt emission \cite{Narayan:1992iy, Usov:1992zd}. This occurs at a distance of around $10^{12} - 10^{14}$~cm from the center. The jet subsequently moves forward and plunges into the external medium surrounding the burster, launching a relativistic shock that generates the afterglow radiation. Along with the relativistic forward shock, for baryonic ejecta, there can also be a reverse shock moving to the ejected material. This reverse-forward shock system is dynamically coupled. Below, we discuss the dynamics of the forward shock. See section-\ref{RS} for more details of the reverse shock. 

\subsection{Dynamics of the forward shock}
\label{FS}
The swept up material, which is accumulated in the downstream of the forward shock, is heated up to relativistic temperatures. Random motion Lorentz factors of protons in the downstream is roughly equivalent to the bulk Lorentz factor of the shock. When the mass equivalent of the downstream thermal energy, $\Gamma_0 m_{\rm sw}$ (where $\Gamma_0$ is the initial bulk Lorentz factor of the outflow and $m_{\rm sw}$ the rest mass of the swept-up material), approaches the ejected mass ($M_{\rm ej} = E_k/\Gamma_0 c^2$, where $E_k$ is the kinetic energy of the ejected material) the shock starts to decelerate. The observer frame time $t_{\rm dec}$ corresponding to the deceleration radius ($r_{\rm dec}$) is $\sim r_{\rm dec}/(2 \Gamma_0^2 c)$. 

In the simplistic form of the afterglow model, the forward shock dynamics is divided into four phases, separated by three observer frame epochs. First one is $t_{\rm dec}$ mentioned above, before which the fireball moves with constant velocity ($\Gamma = \Gamma_0$). In the post-deceleration relativistic phase, $\Gamma(r)$ follows the self-similar Blandford - McKee dynamics \cite{Blandford:1976uq}, where $\Gamma(r) \propto r^{-g}$  and the value of the scaling index $g$ is decided by the radial density profile of the ambient medium and the downstream energy loss rate (adiabatic vs radiative fireball). The third phase beginning with the jet break time $t_j$ is explained below. The fireball finally evolves into the non-relativistic phase at $t_{\rm nr}$ and asymptotically attains the self-similar Sedov-Taylor dynamics. 

\subsubsection{Jet effects in the afterglow dynamics}
\label{jet}
As the shock decelerates, the bulk Lorentz factor $\Gamma (r)$ drops down and at some point it becomes comparable to $1/\theta_j$, where $\theta_j$ is the initial half-opening angle of the jet. According to simple analytical models \cite{Rhoads:1999wm}, dynamical effects due to the lateral spreading of the jet becomes dominant roughly at this epoch. At the corresponding observer frame time, called the jet-break time ($t_j$), afterglow lightcurve decay is expected to steepen achromatically. Numerical hydrodynamic simulations by \cite{vanEerten:2010mm} found that (i) a logarithmic spreading to be more appropriate (i.e., a nearly non-spreading jet) and (ii) jet break is chromatic in radio frequencies which are optically thick. Further simulations by \cite{Wygoda:2011vu} on the other hand found lateral expansion is important when $\Gamma$ goes below $1/\theta_j$, but only for narrow jets (jets with $\theta_j < \sim 12\deg$).

It is to be pointed out that even if there are no dynamical effects (i.e., a non-spreading jet), jet signature appears in afterglow lightcurves through relativistic effects. Due to relativistic beaming, the observer only sees a $1/\Gamma$ cone of the jet. When $1/\Gamma > \theta_j$, i.e., when $\Gamma$ goes below $1/\theta_j$, the observer starts to see the edge of the jet. The lightcurve decay rate increases even in this case, but the steepening is of a lesser magnitude than that of a spreading jet where dynamics also changes.  

However, data of many well observed afterglows are not consistent with the predicted sharp achromatic break from analytical calculations even at higher frequencies \cite{Liang:2007rn, Racusin:2008bx}. Though part of the reason could be poor sampling, extensive broad-band modelling of well sampled lightcurves is required to understand the dynamics of GRB jets. 

The three observer-frame epochs,  deceleration time ($t_{\rm dec}$), jet break time ($t_j$), and non-relativistic transition time ($t_{\rm nr}$), demarcating the four phases of fireball dynamics depend only on the isotropic kinetic energy $E_k$, ambient density, and the initial half-opening angle $\theta_j$ of the jet. Additional parameters of the problem appear from synchrotron physics. 
 
 
\subsection{Afterglow spectrum}
\label{spec}
The afterglow spectrum emanates from the shock downstream, where electrons are accelerated to a non-thermal power-law distribution, and magnetic field is enhanced. These electrons radiate predominantly via the synchrotron process. The synchrotron flux depends on the fraction $\epsilon_e$ of the shock thermal energy in power-law electrons, the fraction $\epsilon_B$ in downstream magnetic field, and the electron energy distribution index $p$. If the fractional energy content in the downstream magnetic field is much higher than that in the electron population, synchrotron self-Compton (SSC) emission may also become important. However, we ignore this process in the rest of this review.

\begin{figure*} 
\begin{center}
\includegraphics[scale=0.3]{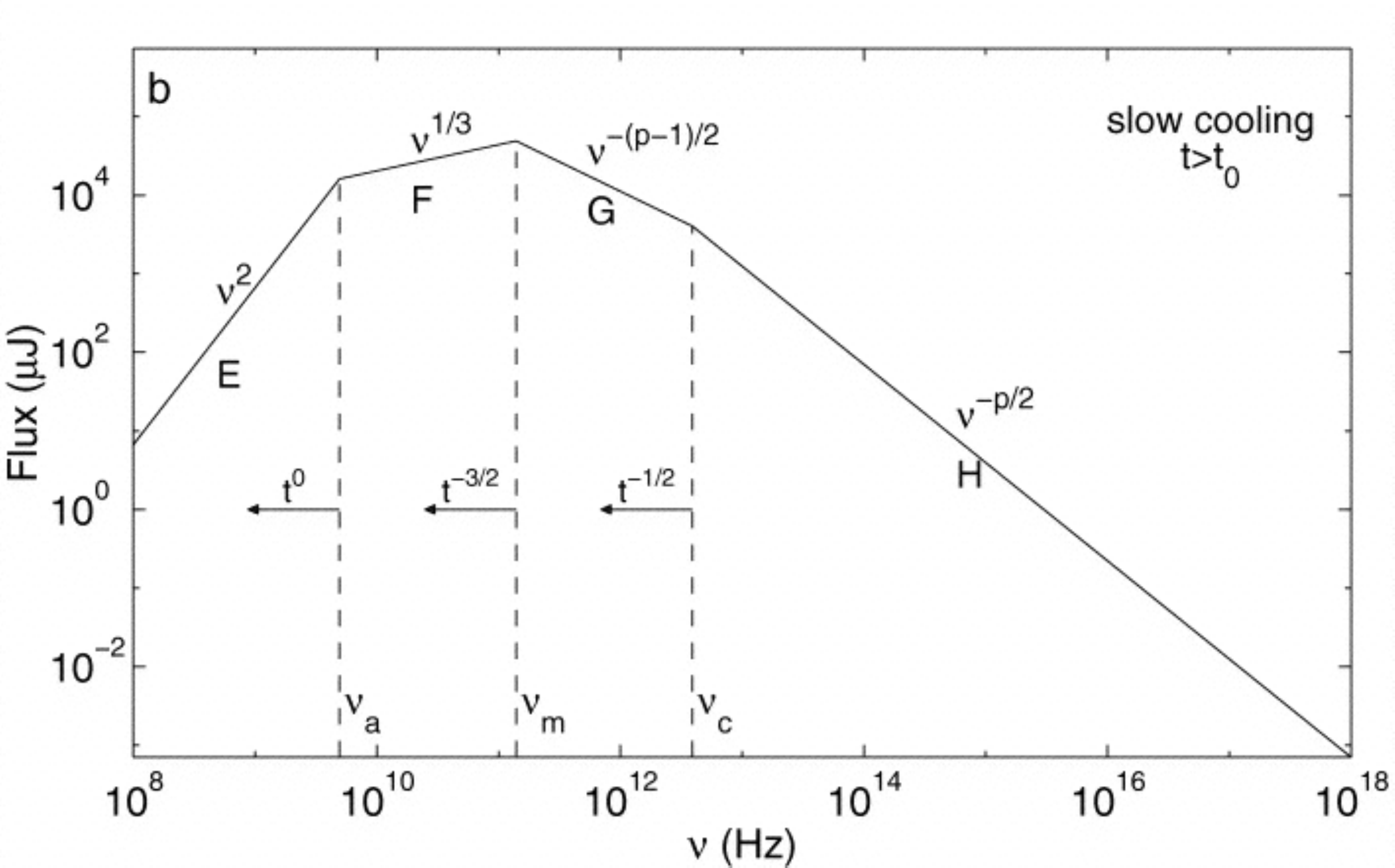}
\includegraphics[scale=0.3]{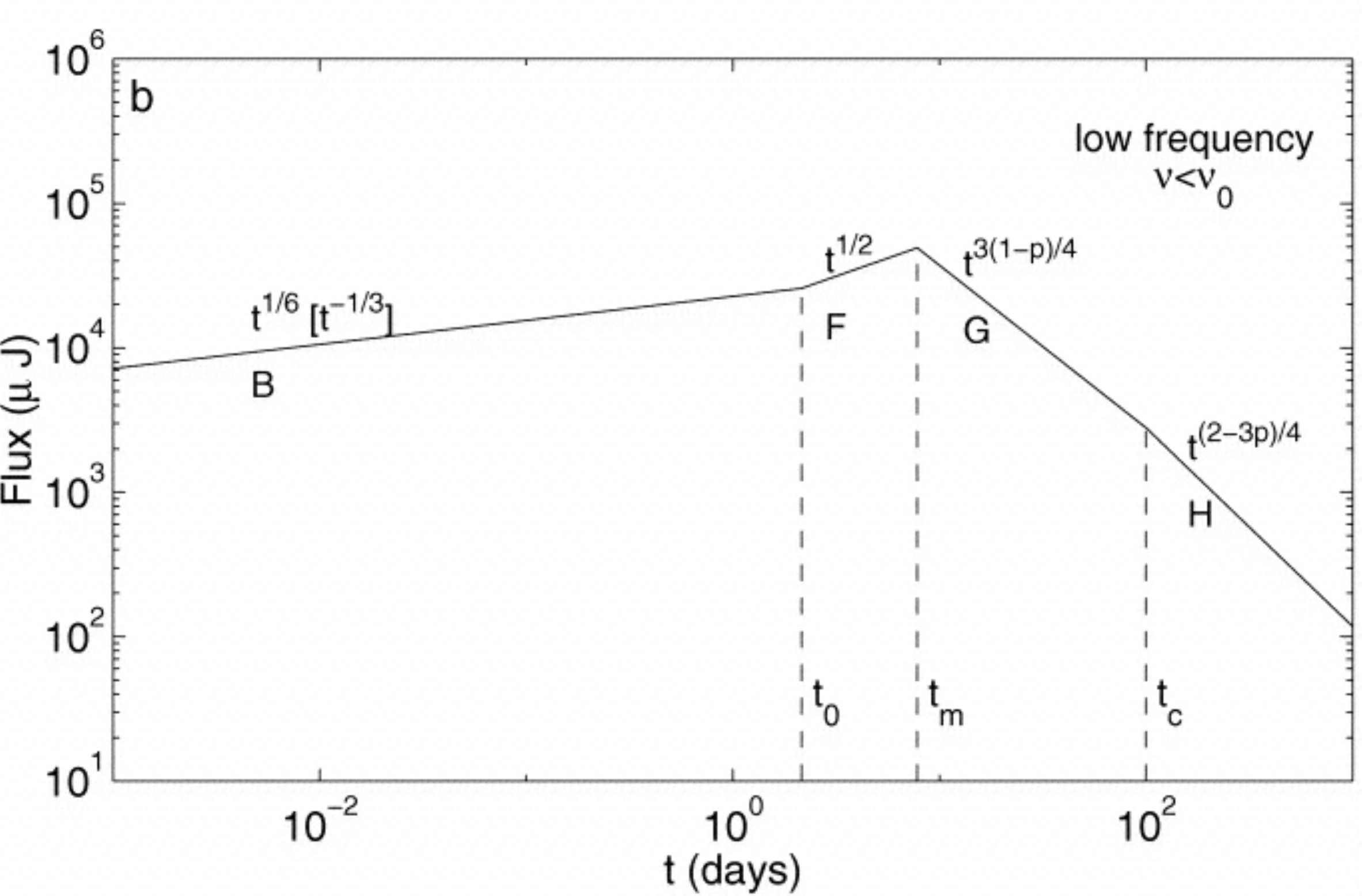}
\caption{Standard fireball model in its bare-bone version. (Left:) The instantaneous synchrotron spectrum and (right: ) low frequency lightcurve. Reproduced from \cite{Sari:1997qe}.}
\label{speclc}
\end{center}
\end{figure*}

The instantaneous synchrotron spectrum is defined by three spectral breaks: (i) frequency $\nu_m$ corresponding to the lowest electron Lorentz factor $\gamma_m$ of the injected electron distribution, (ii) the cooling break $\nu_c$ corresponding to the electron Lorentz factor $\gamma_c$ above which radiative losses are severe, and (iii) the self-absorption frequency $\nu_a$ below which the fireball is optically thick to synchrotron self-absorption. The fourth and final parameter characterizing the instantaneous spectrum is the flux normalization $f_m$. This is the spectral peak at a given time, and it appears at min($\gamma_m, \gamma_c$). Readers interested in deriving these spectral parameters can refer \cite{Wijers:1998st}. Figure-\ref{speclc} (left) gives an example spectrum. There is a unique, but coupled, mapping between the spectral parameters $\nu_m, \nu_c, \nu_a$, $f_m$ and the physical parameters $E_k, n, \epsilon_e, \epsilon_B$ (each spectral parameter depends on at least $3$ if not all of the physical parameters). Hence, broad-band modelling is necessary to faithfully infer the physical parameters of the emission region (in the ideal condition, the spectral index $p$ can be measured directly from the spectrum, however this can often get complicated due to poor data sampling and absorption due to the host galaxy gas and dust column). 

As the shock slows down, downstream thermal energy density goes down. All four spectral parameters mentioned above, characterising the instantaneous spectrum, depend on the downstream thermal energy density. Therefore, the synchrotron spectrum evolves in time resulting in a time varying flux in all wavelengths (see figure-\ref{speclc} (right)).  Generally, the flux at a given observer frame time $t$ and a given frequency $\nu$ can be written as $f(\nu, t) \propto \nu^{-\beta}  t^{-\alpha}$, where $\beta$ is decided by the particular spectral regime $\nu$ falls in and $\alpha$ depends on both the spectral as well as dynamical regimes discussed above. See this recent review \cite{Kumar:2014upa} for a detailed description of the theoretical framework of the fireball model.

Emission from the reverse shock is seen in the early optical and radio afterglows. We will discuss this next.

\subsection{Reverse shock emission}
\label{RS}
\begin{figure} 
\begin{center}
\includegraphics[scale=0.4]{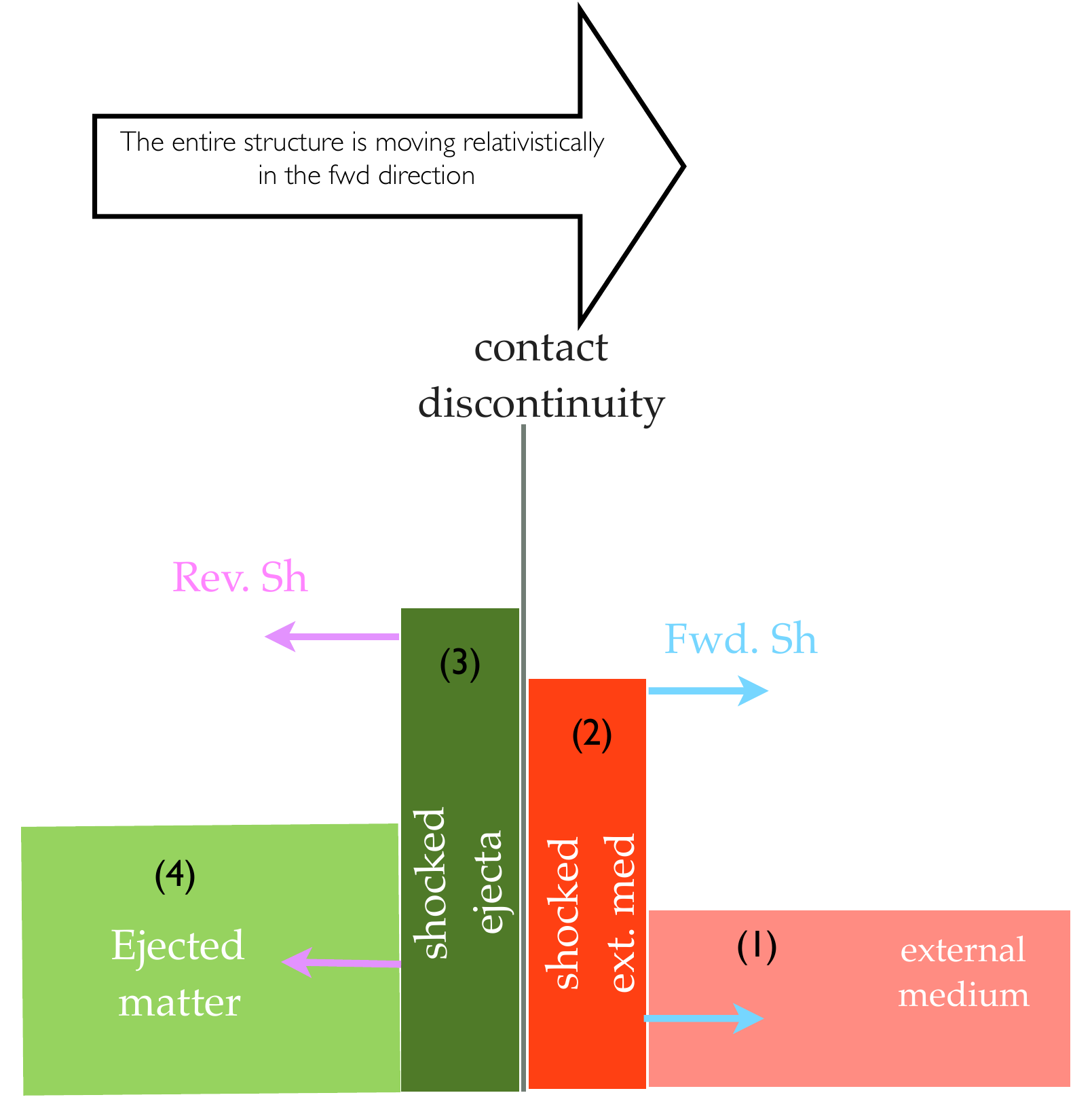}
\caption{The reverse forward shock system, separated by a contact discontinuity (CD). Arrows indicating directions of the shocks are w.r.t to the rest-frame of the CD. With respect to the observer, the entire system is moving towards her in relativistic velocities. Differing heights of each region represents the co-moving number densities.}
\label{sketch}
\end{center}
\end{figure}

Apart from the forward shock, a reverse shock also forms if the ejecta from the central engine is baryonic. The reverse shock (RS) moves into the ejecta itself. The shocked ejecta and shocked ambient medium are separated by a contact discontinuity (see figure-\ref{sketch} for a sketch). Dynamics of both shocks are coupled. The reverse shock starts off with non-relativistic velocities and turns mildly relativistic if the ejected shell is thick enough. Simplistic calculations hence consider RS dynamics in two asymptotic regimes: Newtonian RS (thin shell) and relativistic RS (thick shell) \cite{Meszaros:1999kv, Sari:1999kj, Kobayashi:2000af}. Thin and thick here refer to the physical width of the ejected shell, not to be confused with self-absorption (or the lack thereof) that makes the RS downstream optically thick (or thin).  Newtonian RS crosses the ejected shell at the deceleration epoch $t_{\rm dec}$ described in section-\ref{FS}. For relativistic RS shock-crossing occurs at $\sim T_{90}$. Dynamics of the Newtonian RS depends on the initial bulk Lorentz factor $\Gamma_0$ of the ejecta, in addition to $E_k$ and number density of the ambient medium. For relativistic RS, one more additional parameter, $T_{90}$, representing the width of the ejected shell, enters the picture. For a detailed review, see \cite{Gao:2015lga}. 
Synchrotron spectrum from the reverse shocked ejecta can be calculated from the properties of the downstream the same way as described for the FS in section-\ref{spec}. Reverse shock emission appears in lower frequencies compared to the forward shock. This is because its upstream, the ejected matter, is denser than the ambient medium, the upstream of the forward shock. Hence  the average kinetic energy of the accelerated electrons is smaller compared to that of the forward shock, and thereby the synchrotron spectral peak shifts to lower frequencies. 

Along with the parameters required to explain RS dynamics ($E_k$, ambient density, and $\Gamma_0$ for Newtonian RS, and additional $T_{90}$ for relativistic RS), three more parameters representing the RS downstream micro-physics will complete the full set required to explain the spectral evolution. These are: the fractional energy in non-thermal electrons and magnetic field, ${\epsilon_e}^{\rm RS}$ and ${\epsilon_B}^{\rm RS}$ respectively, and the electron index $p$ in the RS downstream. Time evolution of the RS downstream thermal energy density, number density, and bulk velocity will be different from that of the FS, making its lightcurves to appear quite different from that of FS. 

Temporal peak of the optically thin reverse shock lightcurve (optical/IR/mm) occurs when the shock crosses the ejected shell \cite{Sari:1999kj}. RS peak can hence be used as a probe of $\Gamma_0$ within this simple framework. Since the ejected material is a shell of finite width, optically thin reverse shock emission does not last for a very long time after the shock crosses the shell. Hence RS emission in optical/IR/mm often appears as a fast decaying component in the early afterglow \cite{Sari:1999iz}. 

Standard RS calculation that we described above assumes pressure equilibrium at the contact discontinuity. However, if the RS is long-lived, this assumption breaks down. See \cite{Uhm:2010js} for a formalism that relaxes this approximation. 

\subsection{Deviations from standard model}
Though the standard model treats the prompt emission (internal dissipation) and afterglow (external dissipation) as two separate phases, in reality the border is diffuse. In the basic standard model described above, afterglow calculations are done assuming a single shot of energy deposition in the external medium. However, early X-ray lightcurves from \emph{Swift} revealed that the reality is far from this. In majority of GRBs the central engine appears to inject additional energy in a continuous manner to the already decelerating external shock for a few hundreds of seconds \cite{Nousek:2005fm}. In addition to that in some cases the early afterglow is influenced by spasmodic late energy supply from the central engine appearing in the form of X-ray flares \cite{Burrows:2005ww}. Several authors have incorporated a continuously powered fireball in theoretical afterglow calculations \cite{ZhangMesz:2001, Panaitescu:2005nq, gulli:2006}.

Another major discrepancy between the standard model and observations is in jet breaks (see section-\ref{jet}). 

\begin{figure}
\begin{center}
\includegraphics[scale=0.4]{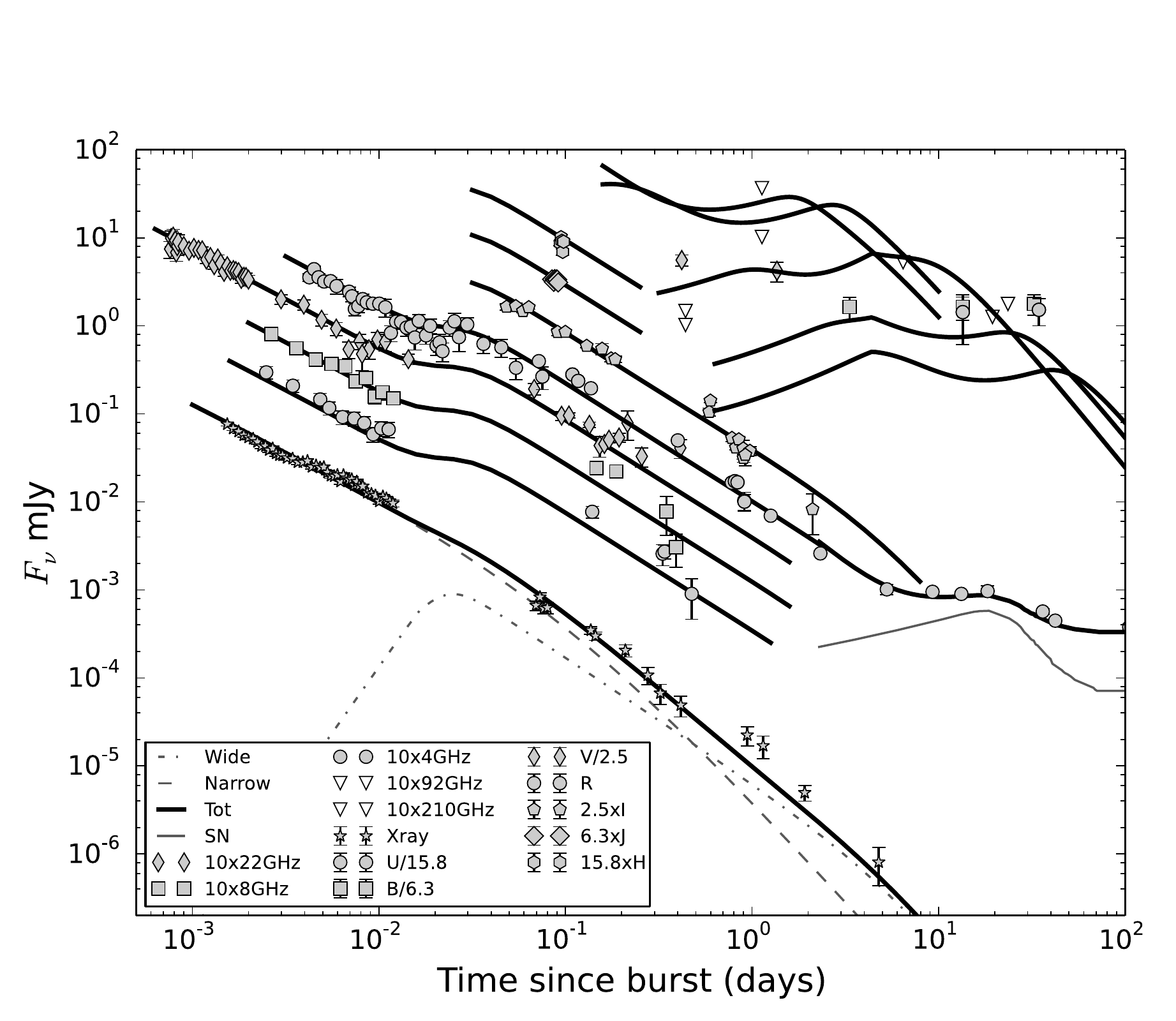}
\caption{An example of broad-band modelling from \cite{Resmi:2012dz}. Here, a two component jet model is used to explain the afterglow evolution.}
\label{0525afig}
\end{center}
\end{figure}

Furthermore, many well observed afterglows show a far more complex evolution than what is described in the standard model. Additional emission components like a wider jet or cocoon are commonly invoked to explain the full evolution of a single well observed burst \cite{Resmi:2012dz} (see figure-\ref{0525afig}). There are also bursts that do not adhere to the simplified version of synchrotron theory and requires various modifications, for example, a time dependent micro-physics of the shock downstream. One good example is the well observed GRB 130427a \cite{Panaitescu:2013pga}.

A uniform top-hat jet, where the jet Lorentz factor and emissivity are uniform within a cone of rigid edges, is assumed in the most commonly used standard model calculations. However, in reality, the jet could be structured with an angular dependence for the above parameters \cite{Meszaros:1997je, Zhang:2001qt}. Afterglows emanating from such a jet definitely will deviate from model predictions assuming a uniform top-hat jet.
%

To summarise, broadband spectral evolution of GRB afterglows can shed light onto the energetics of the explosion, nature of the ambient medium, structure of the jet, and the micro-physics of the shocks. Though a complete understanding requires dense sampling of the multi-wavelength data, in the next section we focus on some of the questions radio afterglows are specifically equipped to address.
\section{Probing GRB physics through associated radio emission}
\label{prob}
First observation of a radio afterglow was in 1997 for GRB970508  by the Very Large Array (VLA) \cite{Frail:1997qf}.  Since then radio observations have played a unique role towards the development of GRB physics. 

We will first discuss one of the important milestones radio observations could achieve - the evidence for relativistic motion. In the remaining three subsections, we will describe three potentially promising avenues to explore with high sensitivity next-generation radio facilities.
\subsection{Inferring relativistic motion through radio observations}
\label{scintil}
As mentioned in section-\ref{basics}, relativistic motion was proposed to resolve the compactness problem. It was radio observations that finally proved the hypothesis. 

The radio afterglow of GRB970508 showed strong (r.m.s variation $\sim 0.8$ mean flux) and rapid ($\delta t < 5$ hr) variability for the initial few weeks \cite{Frail:1997qf}. These fluctuations were also chromatic, i.e., uncorrelated between $4.8$ and $8.4$~GHz radio frequencies. The chromaticity can be seen through the variation of the spectral index $\beta$ of (where the flux $f_{\nu} \propto \nu^{\beta}$) between $4.8$ and $8.4$~GHz frequencies. Such rapid, strong and chromatic variability evidently points to diffractive scintillation, which occurs when rays are diffracted by small scale irregularities of the ionized medium they travel through \cite{Goodman:1997yf}. These rays interfere and cause strong flux modulations. But this occurs only when the source angular size is smaller than the characteristic size of the irregularities in the local ISM. Hence, only very compact sources show diffractive scintillation. 

\begin{figure*}
\begin{center}
\includegraphics[scale=0.22,angle=270]{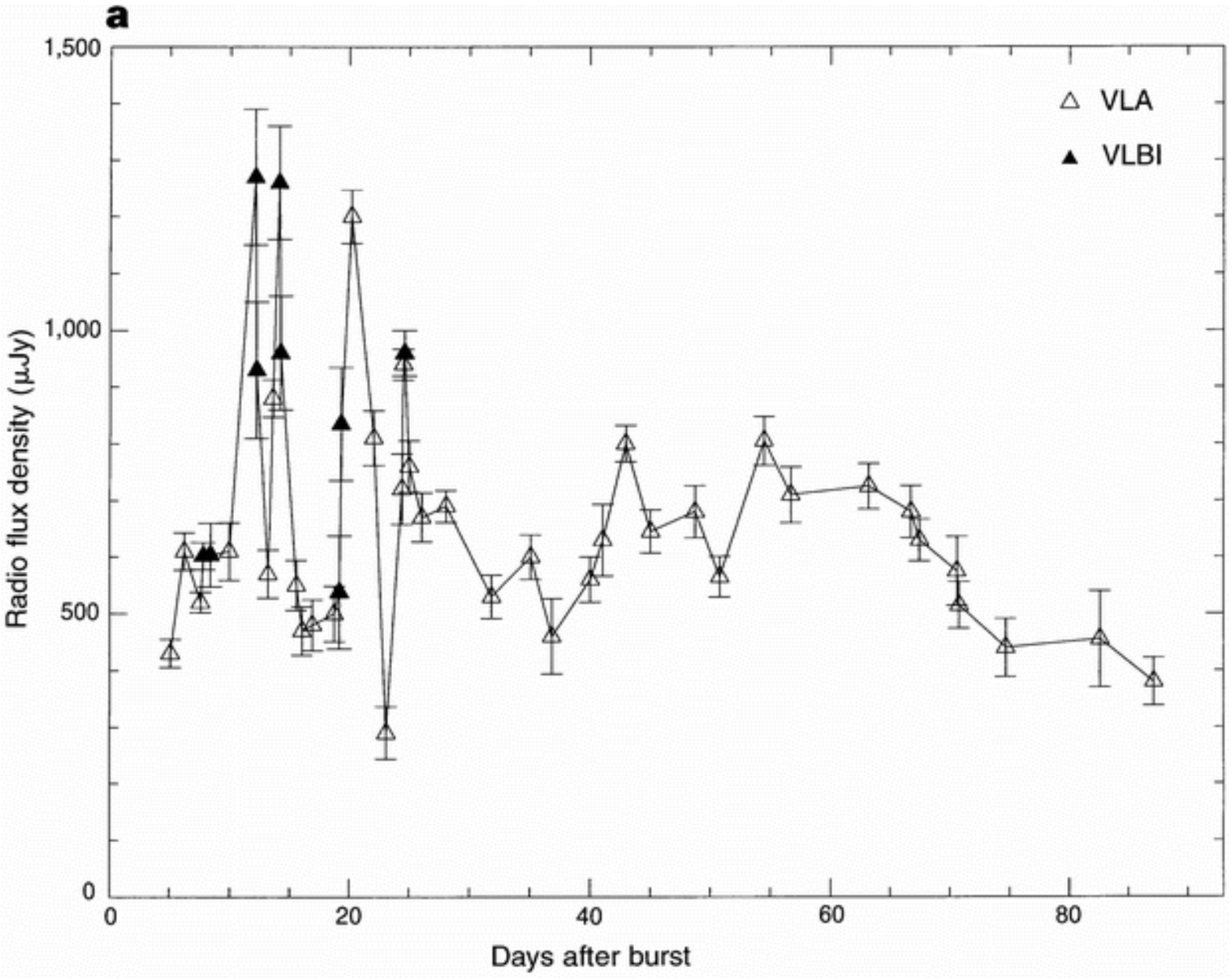}
\includegraphics[scale=0.22,angle=270.]{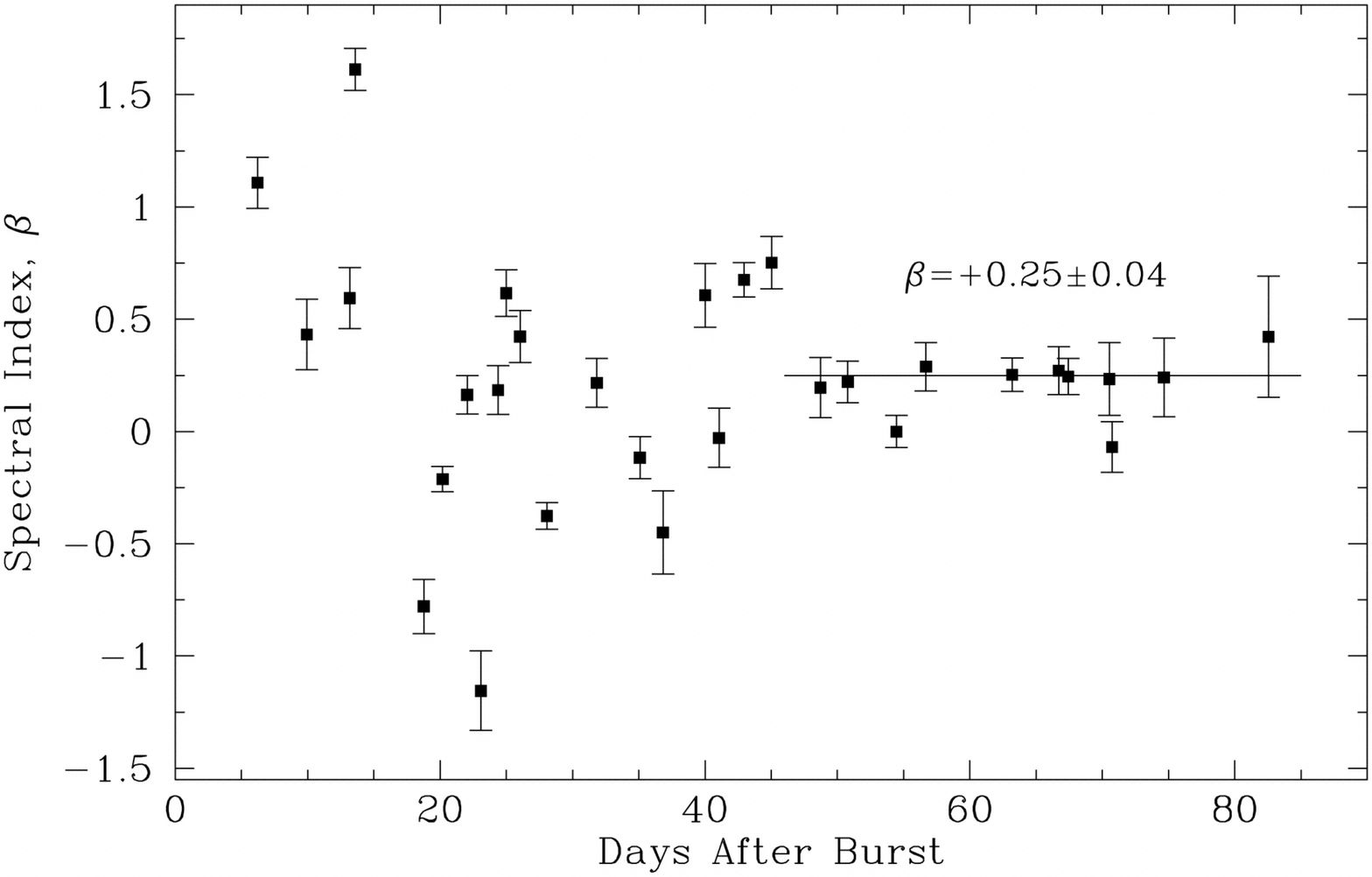}
\caption{The scintillating radio afterglow of GRB 970508. (left) Early ($< 30$~day) oscillations in $8$~GHz flux indicating diffractive scintillation, that switches off once the source size crosses a threshold (reproduced from \cite{Frail:1997qf}). (right) The $4 -- 8$~GHz spectrum of the afterglow (reproduced from \cite{Frail:1999hk}). We can see the spectral index variation drastically reduces around $40$ days, confirming the diffractive scintillation hypothesis.}
\label{scintifig}
\end{center}
\end{figure*}

After around $40$ days, the variability died down for GRB9705058 and the afterglow began to follow a smooth decay, indicating an increase in source size beyond the threshold for diffractive scintillation by Milky Way's ISM. This served as an evidence of the relativistic motion proposed to resolve the compactness problem (see figure-\ref{scintifig}).

\begin{figure}
\begin{center}
\includegraphics[scale=0.3]{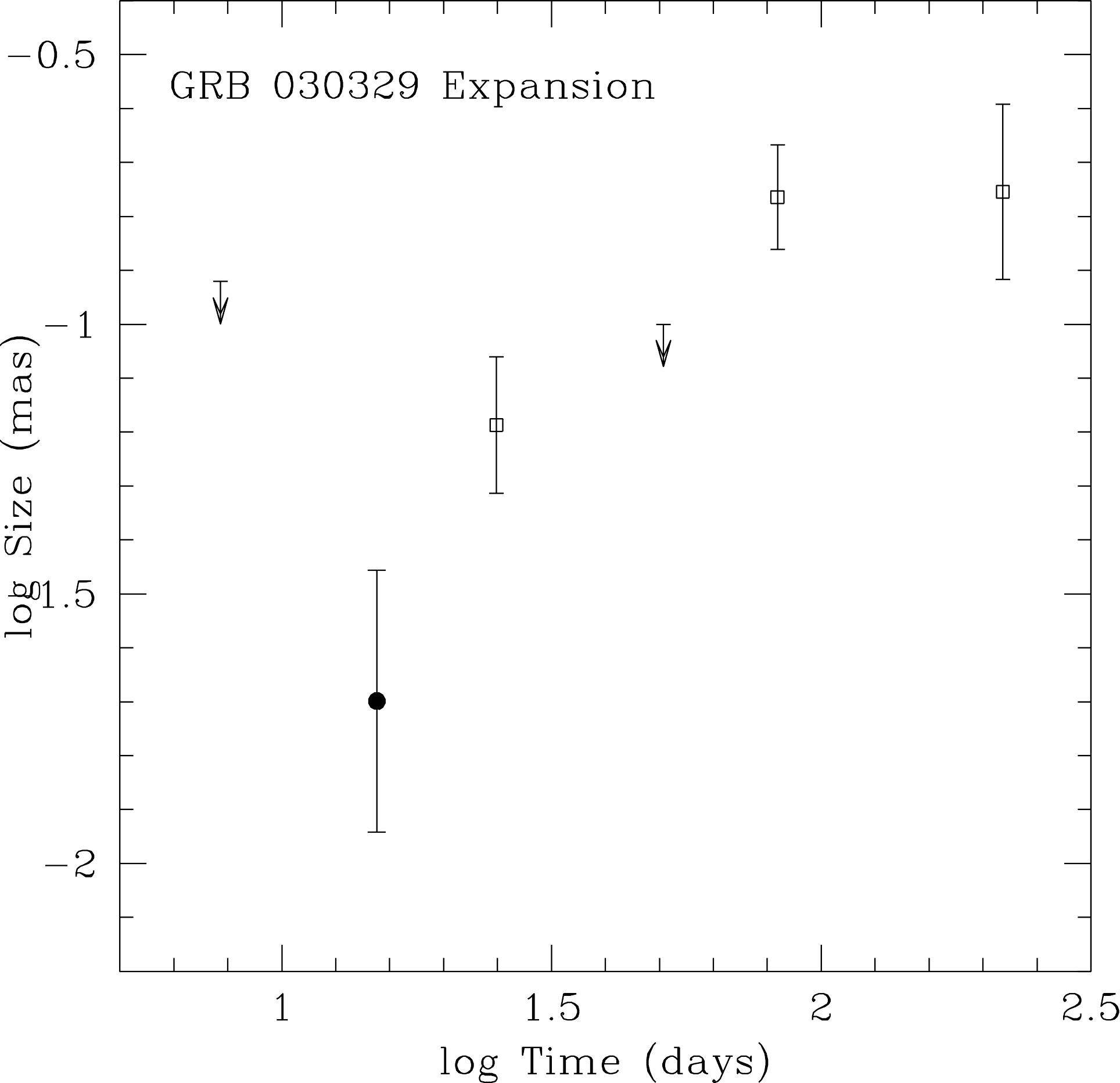}
\caption{Angular size evolution of the radio afterglow of GRB 030329 from VLBI. Reproduced from \cite{Taylor:2004ru}. }
\label{vlbifig}
\end{center}
\end{figure}

However, it was only a matter of time that a direct observation of relativistic motion was possible, once again in the radio band. For the bright nearby ($z = 0.16$, at a luminosity distance of $580$~Mpc) GRB030329, angular size evolution could be obtained through Very Large Baseline Interferometry (VLBI) and tested against models \cite{Taylor:2004wd} (see figure-\ref{vlbifig}). This once again confirmed relativistic bulk motion in GRBs. However, VLBI studies are possible only for bright afterglows.

Compared to afterglow detections in X-ray and optical frequencies, radio afterglow detection statistics is the lowest \cite{Chandra:2011fp}. Upcoming high sensitive radio facilities, including the currently operational JVLA and upgraded-GMRT, can increase the number of radio detections. 
\subsection{Radio reverse shock emission: ejecta magnetization and ambient medium}
Quick and sensitive follow-up observations of the JVLA has recently enabled detection of bright early radio emission, for example from GRB 130427a and GRB160509a, possibly from the reverse shock. 

\begin{figure}
\begin{center}
\includegraphics[scale=0.3]{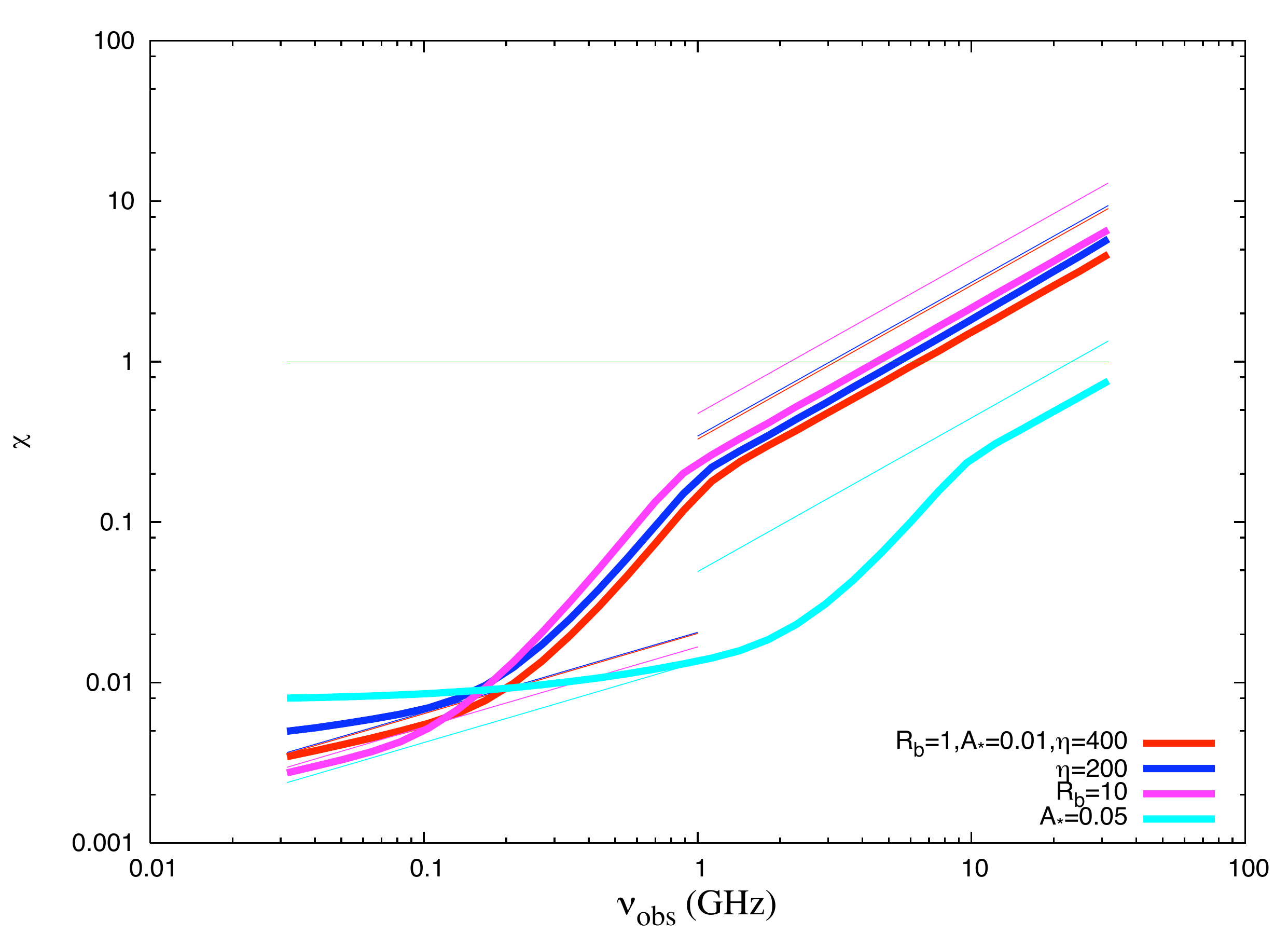}
\caption{Ratio $\chi$ of reverse to forward shock flux at the time of the reverse shock peak, as a function of the observing frequency. RS is very unlikely to dominate in low radio frequencies. Different colors represent different set of physical parameters. Notice the sensitivity of $\chi$ to ambient medium density. A wind driven density profile is assumed in this calculation. Reproduced from \cite{Resmi:2016lmn}.}
\label{RSfig2}
\end{center}
\end{figure}

In radio frequencies, since self-absorption effects are important, characteristics of the reverse shock lightcurve are different from that of the optically thin lightcurve. \cite{Resmi:2016lmn} calculated the self-absorbed RS emission and estimated the detectability of the RS component as a function of varying physical parameters. They found that low frequency radio reverse shock emission is enhanced by low ambient density (see figure-\ref{RSfig2}). This is a consequence of reduced optical depth in the RS downstream due to increased ejecta spreading. For low ambient densities, delayed shock-crossing gives sufficient time for the ejecta to spread and rarefy itself. 

\begin{figure*}
\begin{center}
\includegraphics[scale=0.4]{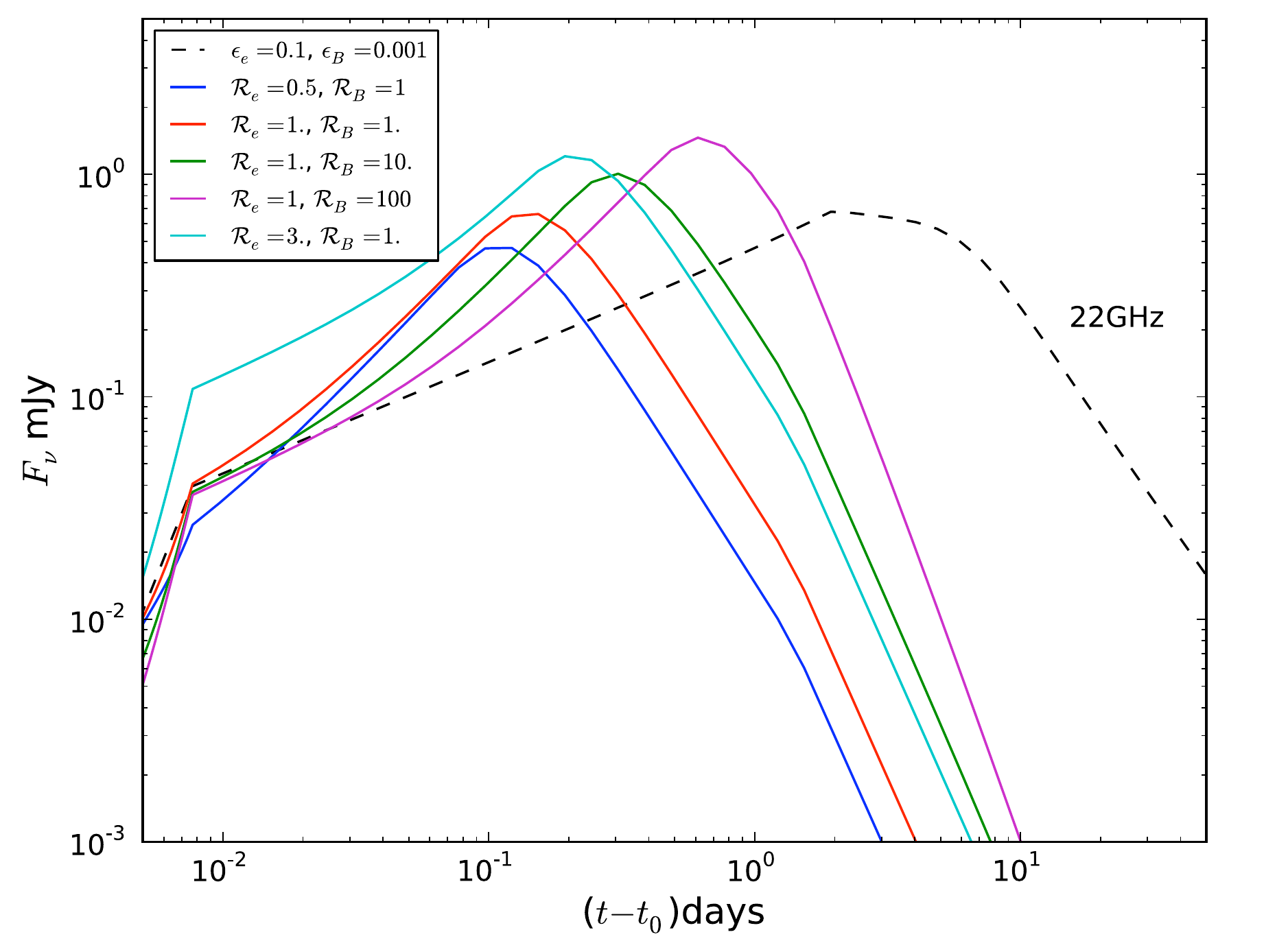}
\includegraphics[scale=0.4]{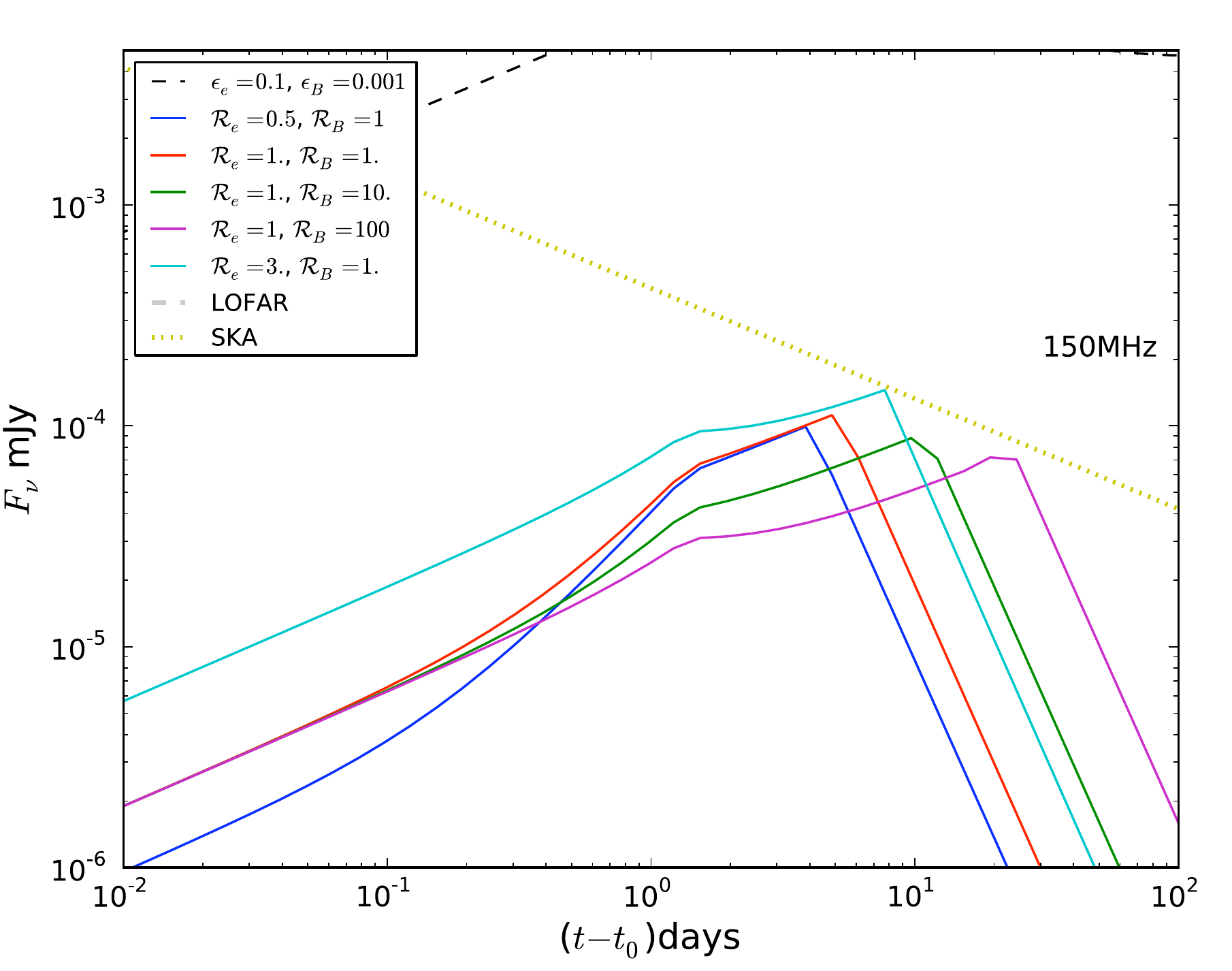}
\includegraphics[scale=0.4]{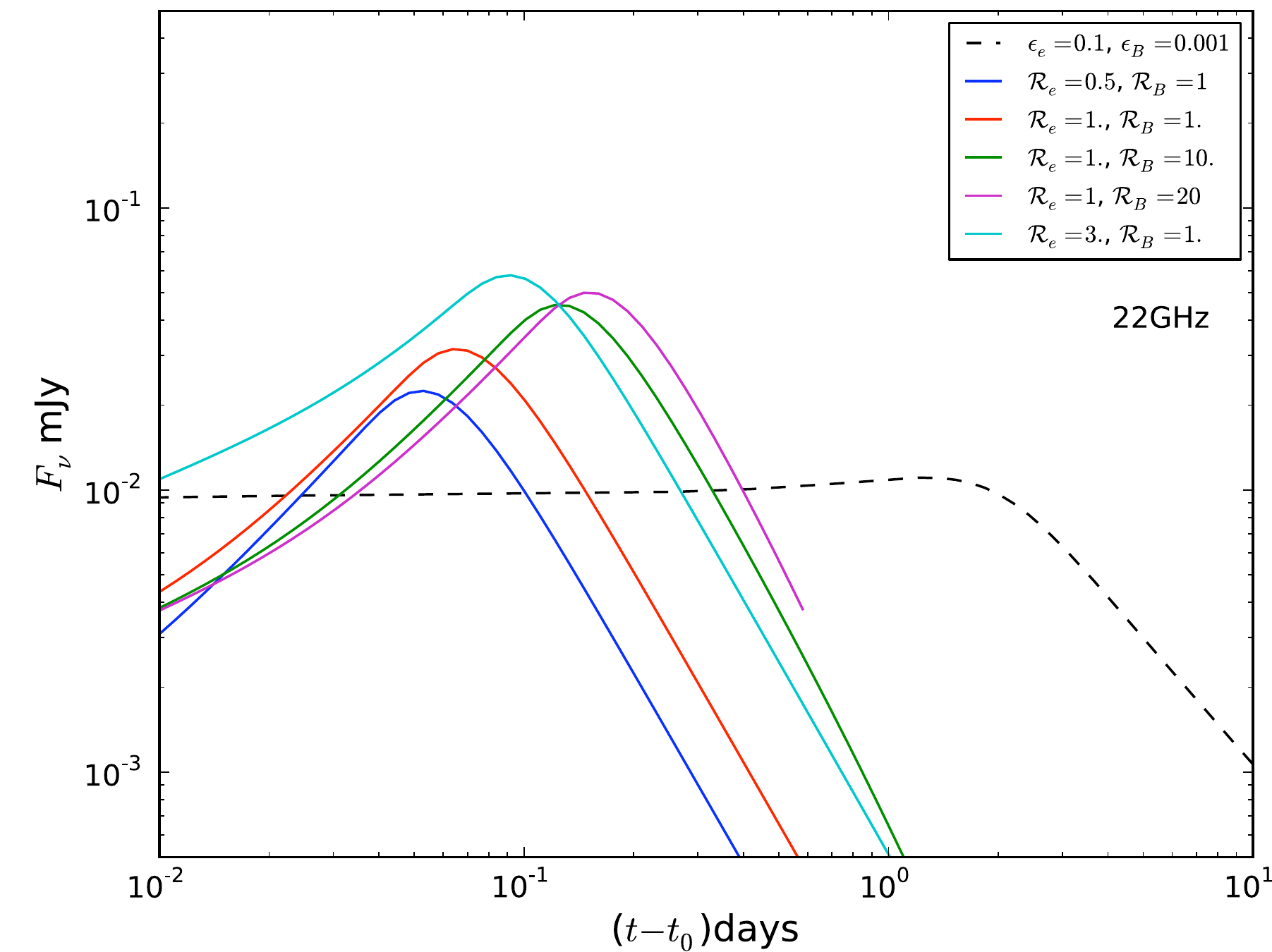}
\includegraphics[scale=0.4]{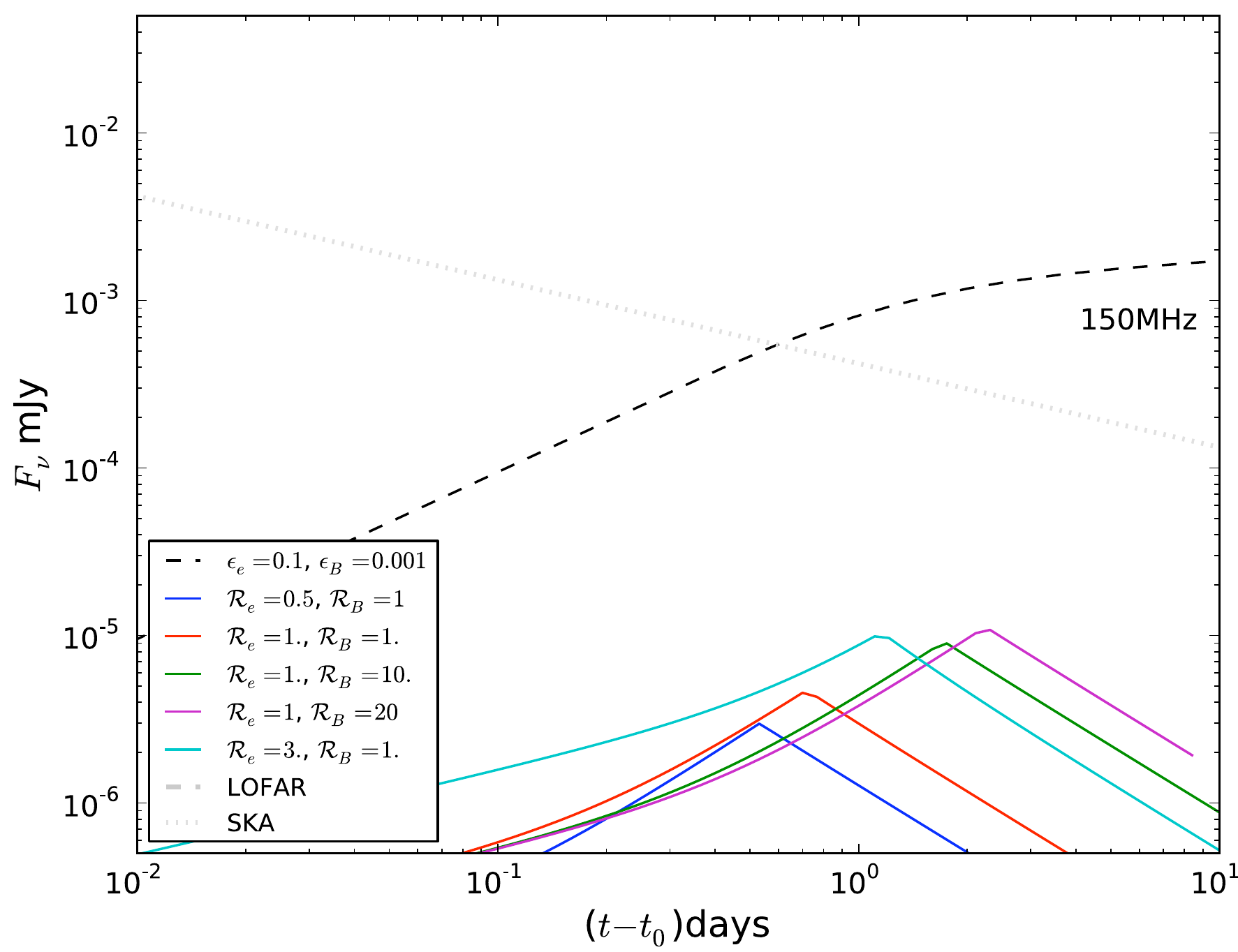}
\caption{Radio reverse shock calculations from \cite{Resmi:2016lmn}. The top panel is for a homogeneous ambient medium and the bottom panel for a wind driven stratified ambient medium. One can see that higher ${\cal R}_B$ delays the reverse shock peak, and that the reverse shock does not dominate in MHz radio frequencies for standard range of parameters.}
\label{RSfig}
\end{center}
\end{figure*}

The reverse shock (RS), for which the ejected material itself is the upstream, is an excellent probe to the central engine, especially its magnetization \cite{Zhang:2004ie, Fan:2004ka}. Moderate magnetization, ${\cal{R}}_B = {\epsilon_B}^{RS}/{\epsilon_B}^{FS} \sim$ $10$ -- $100$, is found to enhance the optical reverse shock emission \cite{Jin:2007uf}. However, increasing ${\cal{R}}_B$ does not increase the radio reverse shock flux, due mostly to self-absorption effects \cite{Resmi:2016lmn}. However, the radio reverse shock lightcurve peak, which occurs when the fireball becomes optically thin, is sensitive to ${\cal{R}}_B$. Ejecta magnetization can be uniquely probed with early optical and radio reverse shock observations. See figure-\ref{RSfig} for radio RS predictions for different parameter sets.

\cite{Resmi:2016lmn} also saw that for typical parameters, unless the ambient medium density is extremely low, the RS never outshines the FS in radio frequencies below $\sim 1 GHz$. 
\subsection{Non-relativistic afterglow}
\label{NR}
\begin{figure}
\label{0329lcfig}
\begin{center}
\includegraphics[scale=1]{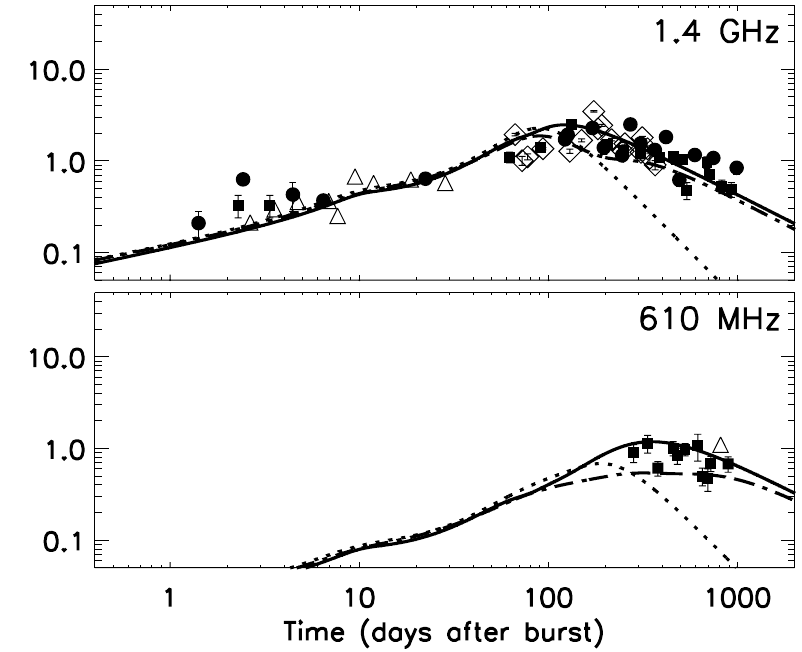}
\caption{Low radio frequency lightcurves of GRB030329 reproduced from \cite{vanderHorst:2007ir}.}
\end{center}
\end{figure}

A singular feature of radio frequencies is that the fireball remains visible for months to years. The text book example is the radio bright GRB030329, which was the first burst observed in MHz radio frequencies. See the radio image of the afterglow around four months since burst by GMRT in figure-\ref{gmrtmap}.  The radio afterglow was observed for more than 5 years with GMRT in $1280$ and $610$~MHz frequencies (see figure-\ref{0329lcfig} for the lightcurve). In these time scales, the shock dynamics changes to non-relativistic velocities, a regime probed in detail only in three GRBs so far \cite{Berger:2004rn, vanderHorst:2007ir}. Single epoch radio observations post $\sim 100$ days are available for around two dozen bursts \cite{Shivvers:2011uj}. Under the assumption that the outflow becomes spherical, energy estimates done in these time-scales are free from the uncertainties in the jet opening angle, unlike broad-band modelling concentrating on the earlier phases. The total (calorimetric) energy in the explosion can be measured and it may allow differentiating between models of central engine of GRBs (magnetar vs black-hole powered) \cite{Wygoda:2011vu}. Thereby, bright radio afterglows provide a unique angle to probe the GRB energetics and central engine. 

However, Calorimetric estimates can be affected by two factors : (i) non-spherical geometry in the outflow, (ii) off-axis observer.
\cite{Wygoda:2011vu} found that the difference in energy estimates due to non-spherical geometry in calorimetry is negligible. However, off-axis observers ($\theta_v \gg \theta_j$, who miss the burst itself) may find calorimetric measurements to be significantly different from the true energetics.
\begin{figure*}
\label{gmrtmap}
\begin{center}
\includegraphics[scale=0.3]{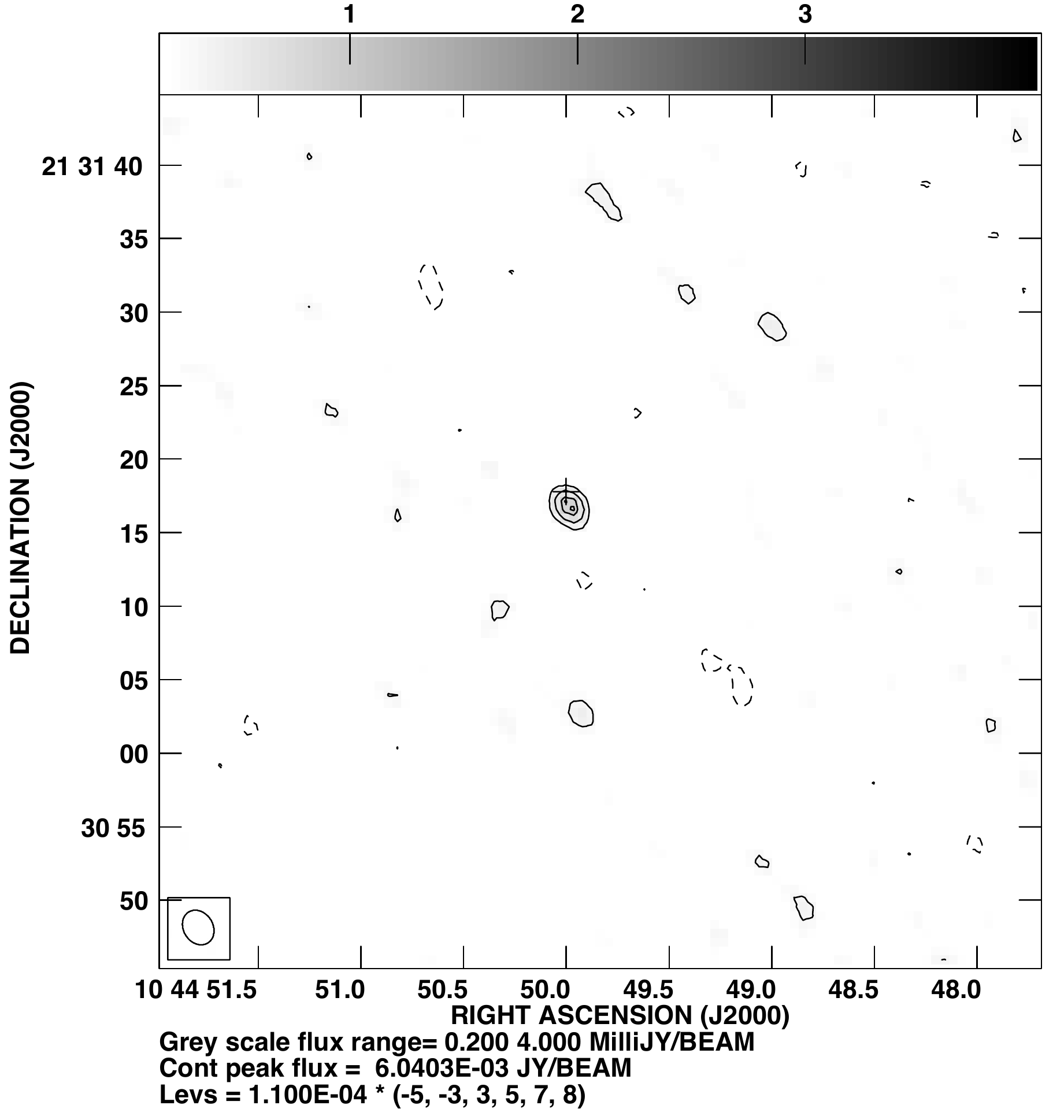}
\includegraphics[scale=0.3]{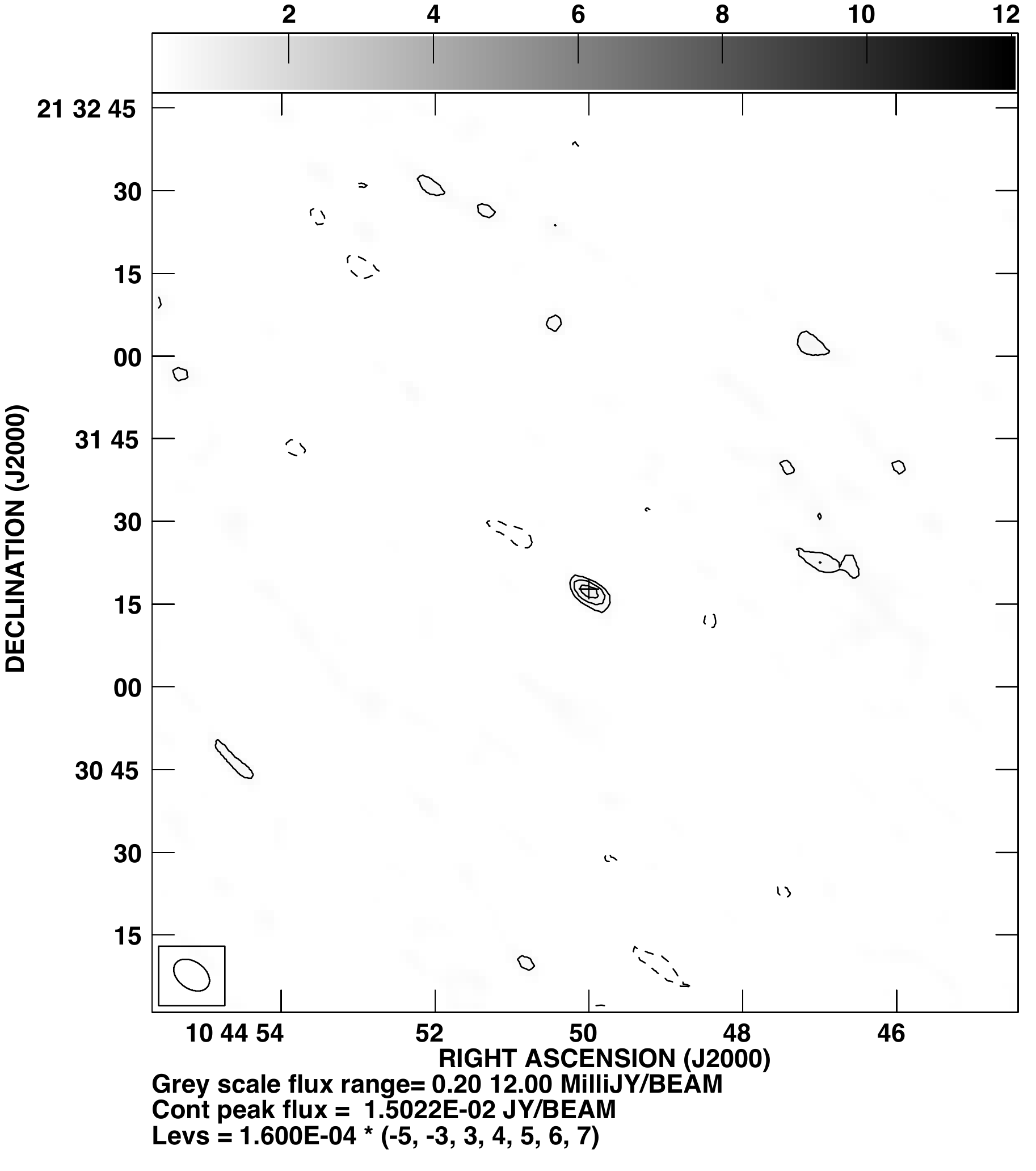}
\caption{Radio continuum maps of GRB 030329 in $1280$MHz (left) and $610$~MHz (right). These observations are at $117$ and $116$ days post the burst respectively.}
\label{gmrtmap}
\end{center}
\end{figure*}

\begin{figure}
\label{boxfitfig}
\begin{center}
\includegraphics[scale=0.3]{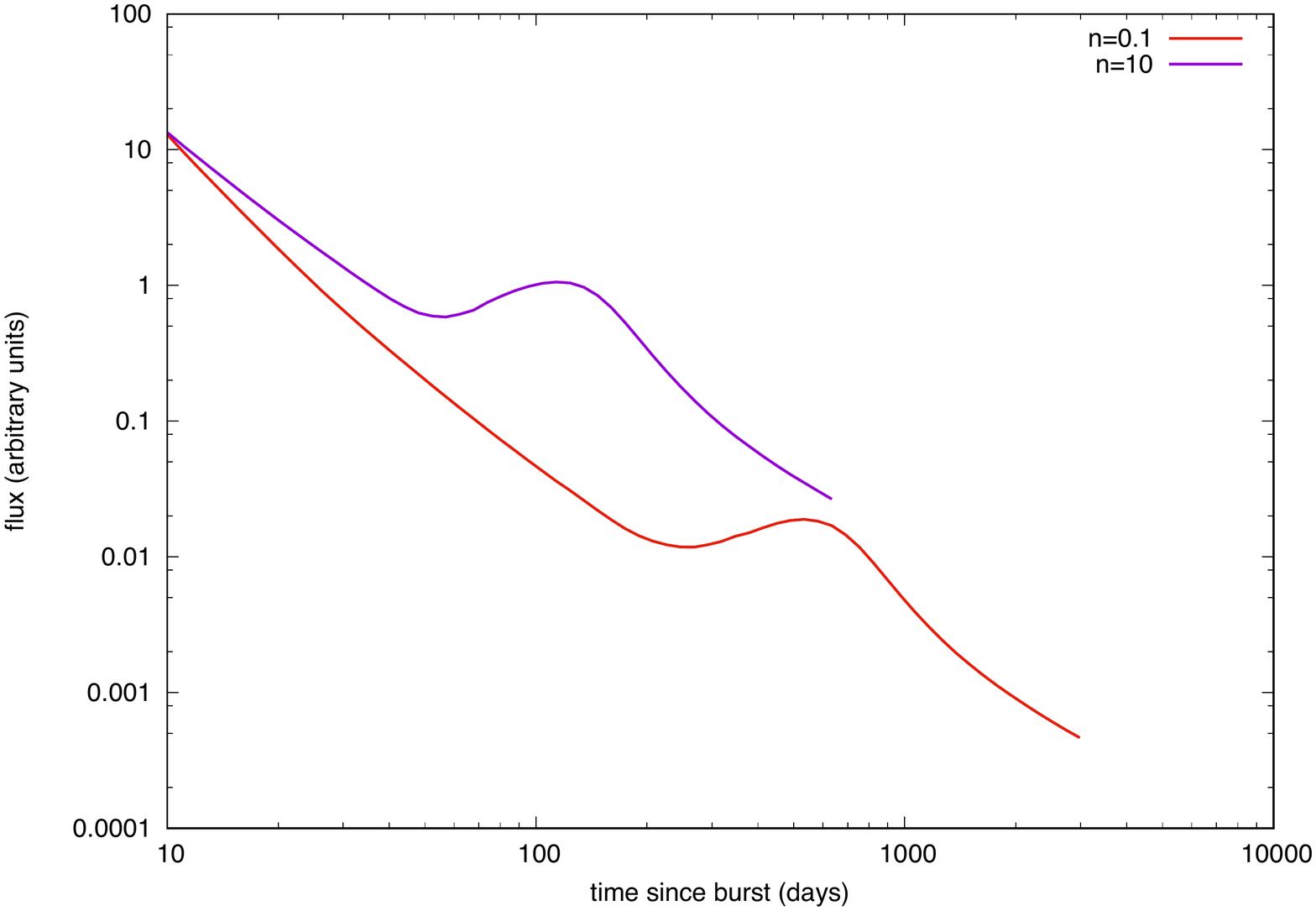}
\caption{Late radio lightcurves ($5$~GHz) for two different densities of the homogenous ambient medium ($0.1$ and $10$~atom/cc) showing the counter-jet bump. Jet opening angle of $\sim 5$ degrees is used. Produced using the publicly available afterglow numerical hydrodynamical code \emph{BOXFIT} \cite{vanEerten:2011yn}.}
\end{center}
\end{figure}

Another feature expected in the non-relativistic phase is the emergence of the counter-jet \cite{Li:2004tu}. Assuming the central engine produces a bipolar relativistic outflow, emission from the jet that moved away from the observer never appears until the relativistic de-boosting is lifted off. As the fireball becomes non-relativistic, photons from the counter-jet (CJ) arrives to the observer after travelling an additional $2 R$ distance (where $R$ is the radius of the fireball at the time of emission) compared to the photons from the forward moving jet. Hence the CJ photons at a given observed time will correspond to an emission time earlier than that of the forward moving jet. Since the jet rest-frame emissivity decreases with time, assuming identical emissivities for both jets, the counter-jet when emerges should be brighter than the forward jet, and appears as a bump above the forward jet lightcurve as per analytical calculations \cite{Li:2004tu}.  There has so far been only one burst, GRB 030329, where the CJ emission could be searched for. However, no conclusive evidence is seen in late time GMRT observations \cite{vanderHorst:2007ir}. However, numerical hydrodynamical simulations show that the CJ emission never emerges above that of the forward jet, however appears as an excess emission for hard-edged jets \cite{cjsimln} \ref{boxfitfig}. In the era of SKA where prolonged radio follow-up will be possible, contribution from the counter-jet should be detectable. Using CJ emission, the ambient medium density structure and the jet structure can be modelled in detail \cite{cjsimln}.

\subsection{Radio emission associated with short GRBs}
\label{short}
All individual bursts described in the above subsections belong to the long GRB class. Though the same afterglow physics applies to short bursts too, there are very few radio afterglow detections from short GRBs \cite{Fong:2015oha}. 

Short bursts have systematically dimmer afterglows in all bands, but from the detection statistics it appears that they are particularly dimmer in radio.  \cite{Fong:2015oha} has a sample of $103$ short bursts observed in the Swift era. X-ray detection statistics is $\sim 69$\%, in optical band it is around $29$\%, and in radio band it is a mere $4$\%. While X-ray and optical detection rates are respectively $1.3$ and $2.5$ times smaller than the total (long \& short) burst sample, the radio statistics of short bursts is an order of magnitude lower. 

Radio flux depends on the jet kinetic energy and ambient medium density along with other parameters (see section-\ref{std}). It is important to note that the short bursts have lower isotropic energies. Moreover, compact object mergers are expected to occur in low density environments, owing to the large systemic velocities of the compact binary system arising from their natal kicks. Thus the faint radio emission from short bursts is in accordance with the popular progenitor model.

\begin{figure}
\label{lastfig}
\begin{center}
\includegraphics[scale=0.45]{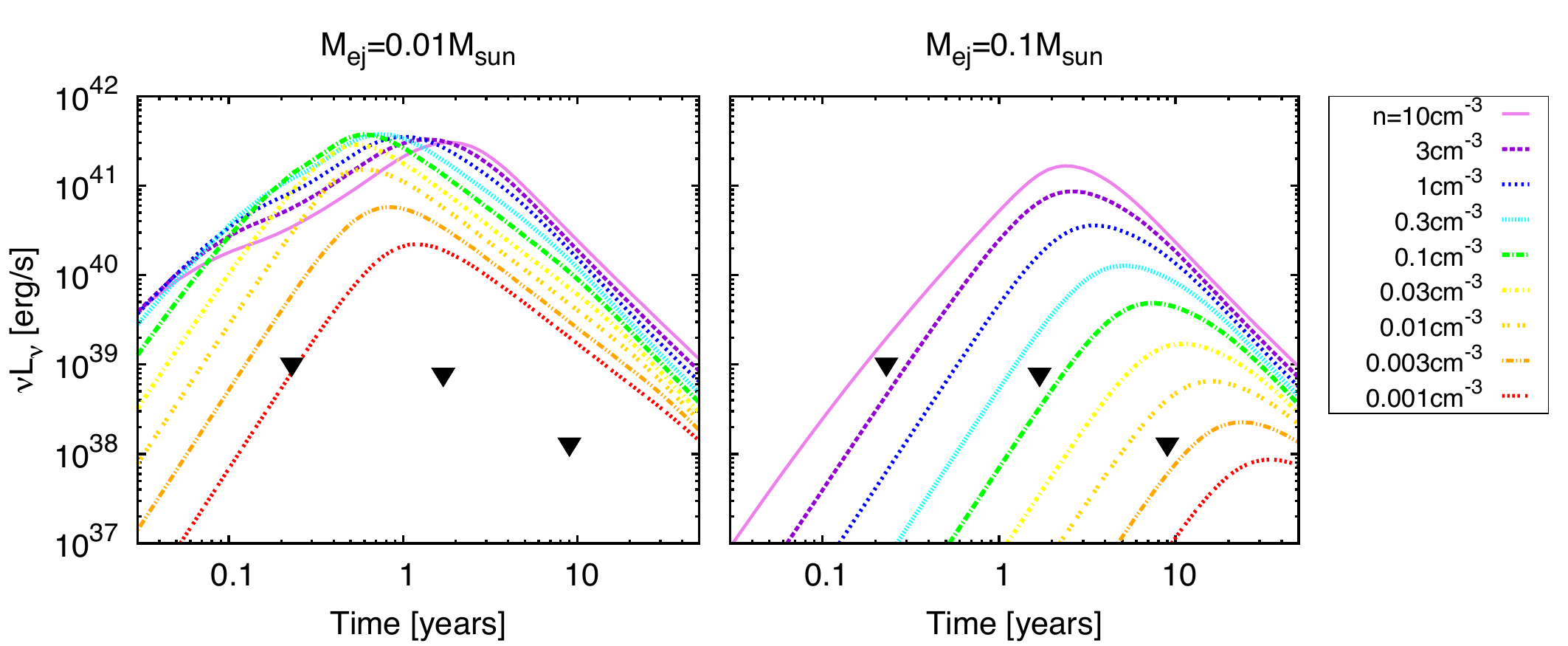}
\caption{Search of non-relativistic merger ejecta from short bursts. Reproduced from \cite{Horesh:2016dah}.}
\end{center}
\end{figure}

An exciting conjecture in the context of short bursts is the predicted radio emission from the tidally disrupted neutron star during the merger \cite{Nakar:2011cw}. The amount of mass tidally ejected will depend on the masses and equations of state of the neutron star. According to numerical simulations of binary mergers, the typical mass in this ejecta is $10^{-2}$ -- $10^{-4} M_\odot$, and the velocities are of the order of $0.2 c$ \cite{Hotokezaka:2012ze}. Since the ejecta is mildly relativistic, the deceleration time, which is the typical timescale where the flux peaks, is of the order of years. Moreover, the synchrotron spectral peak will also be in MHz frequencies \cite{Nakar:2011cw}. As the ejecta is nearly spherically symmetric, detection possibilities are better than the jet emission (short GRB) provided the radio signal is bright enough. 
There have been a few searches for this emission using the VLA and ATCA \ref{lastfig} \cite{Metzger:2013cka, Horesh:2016dah}.  Unless the ambient density is high enough, current radio facilities are unlikely to detect this emission. A more optimistic possibility is if the merger ejecta is further energized by a central magnetar created from the merger. Thus, upperlimits are used to draw constraints on the presence of a magnetar at the central engine and its energy input to the merger ejecta \cite{Metzger:2013cka, Horesh:2016dah}. Again, high sensitive future instruments operating in lower frequencies has the potential to detect this emission. 
\section{Possible diversity in radio afterglows?}
\label{q}
From the statistics of radio observations by VLA till 2011, \cite{Chandra:2011fp} has found that the rate of radio afterglow detections amounts to only 30\% while that of X-ray and optical are around 90\% and 75\% respectively. The lower detection rate of radio AGs is typically attributed to poor detection sensitivity and lack of quick follow up observations \cite{Chandra:2011fp}. The Arcminute Microkelvin Imager Large Array (AMI-LA) operational since 2016 has a quick response but perhaps due to its poor sensitivity, only 15\% of the triggers it responded to so far has resulted in radio detection (statistics from AMI website). 

However, there are claims that the low statistics of radio afterglows does not belong to instrumental effects alone but has its roots in an intrinsic diversity in radio afterglows. The conjecture is of two classes of bursts: radio-bright and radio-faint. \cite{Hancock:2013ena} did a visibility stacking analysis using $737$ VLA observations of $178$ GRBs and found that the radio-faint GRBs are not the tail of a single distribution but are a second population less luminous in all wavelengths. They also found that the radio bright GRBs have typically large isotropic energy in $\gamma$-rays.  \cite{Lloyd-Ronning:2016jfr} observed a correlation between radio afterglow signatures and duration of the prompt pulse. According to them, the radio-quiet bursts have intrinsically shorter prompt emission. Both authors hypothesise that the difference is of physical origin. \cite{Hancock:2013ena} suggest that the radio-faint population results from a different central engine or is associated with a different prompt emission mechanism compared to the radio-loud sample. In almost similar lines, \cite{Lloyd-Ronning:2016jfr} suggests that the diversity perhaps is related to burst progenitors. Though an interesting conjecture, this proposed diversity needs to be explored further in detail through deep follow up using high sensitive radio instruments and thorough modelling efforts.
\section{Future of radio afterglow studies}
\label{future}
Though we have studied a large number of Gamma Ray Bursts and afterglows, there are several fundamental questions yet to be fully answered. For example, how collimated is the outflow launched by the central engine? Jet collimation bears important significance in the calculation of the true rate of GRBs and the total energy budget in the explosion. The total energy budget is important in differentiating between different central engine candidates. The leading models for the same are a black-hole torus system and a highly magnetized rapidly spinning neutron star. It is still not clear which is more favoured, or which one of this could be the potential candidate for a given burst. One of the ways to probe this is by studying the early optical and radio afterglows which may carry signatures of the magnetization of the central engine. Another option is to do calorimetry on late afterglow, seen only in radio band, and obtain the energetics of the burst without involving the jet collimation. Both have increasing potential now in the light of upcoming new generation radio telescopes.

Due to the low number of deep and long radio observations, we are yet to know if there exists any major deviations from the standard picture. \cite{Frail:2003nv} has indicated such a possibility. To explore further, radio detection statistics needs to be improved, which is expected in the SKA era.  

Emission from merger ejecta expected to be detected in lower radio frequencies will be a unique electromagnetic counterpart to binary compact object mergers, apart from GRB jet emission.

Radio band is the lesser explored part of the Gamma Ray Burst counterpart spectrum. In near future, this picture is going to change and a new set of insights and challenges are expected to emerge.

\appendix

\section*{Acknowledgement}
The author thanks the serene atmosphere and the friendly staff at the ``floating cafe'' of Veli tourist village, Trivandrum, where this draft took shape over several cups of afternoon tea. Development of the Boxfit code was supported in part by NASA through grant NNX10AF62G issued through the Astrophysics Theory Program and by the NSF through grant AST-1009863. Simulations for BOXFIT version 2 have been carried out in part on the computing facilities of the Computational Center for Particle and Astrophysics (C2PAP) of the research cooperation "Excellence Cluster Universe" in Garching, Germany. The author acknowledges collaborations with D Bhattacharya, K Misra, and B Zhang.

%
%
%
%


\bibliographystyle{mnras}
   \bibliography{ref}
\end{document}